\documentclass[aps,prl,twocolumn,preprintnumbers,groupedaddress]{revtex4}

\usepackage{amssymb}
\usepackage{epsfig} 
\usepackage{amsmath,color} 
\usepackage{graphicx} 

\usepackage{wrapfig}

\newcommand{\ie}{{\it i.e.,\ }}

\newcommand{\beq}{\begin{equation}}
\newcommand{\eeq}{\end{equation}}
\newcommand{\bea}{\begin{eqnarray}}
\newcommand{\eea}{\end{eqnarray}}
\newcommand{\ba}{\begin{array}}
\newcommand{\ea}{\end{array}}
\newcommand{\bi}{\begin{itemize}}
\newcommand{\ei}{\end{itemize}}
\newcommand{\bn}{\begin{enumerate}}
\newcommand{\en}{\end{enumerate}}
\newcommand{\bc}{\begin{center}}
\newcommand{\ec}{\end{center}}
\renewcommand{\l}{\left}
\renewcommand{\r}{\right}

\newcommand{\eq}[1]{Eq.~(\ref{#1})}

\newcommand{\eqss}[3]{Eqs.~(\ref{#1}), (\ref{#2}) and (\ref{#3})}

\newcommand{\GeV}{\mathinner{\mathrm{GeV}}}

\begin{document}

\preprint{FTUV-14-12-25}
\preprint{IFIC-14-79}

\title{Spiral Inflation}


\author{Gabriela Barenboim}
\email[]{Gabriela.Barenboim@uv.es}
\author{Wan-Il Park}
\email[]{Wanil.Park@uv.es}
\affiliation{
Departament de F\'isica Te\`orica and IFIC, Universitat de Val\`encia-CSIC, E-46100, Burjassot, Spain}


\date{\today}

\begin{abstract}
We propose a novel scenario of primordial inflation in which the inflaton goes through a spiral motion starting from around the top of a symmetry breaking potential.
We show that, even though inflation takes place for a field value much smaller than Planck scale, it is possible to obtain relatively large tensor to scalar ratio ($r \sim 0.1$) without fine tuning.
The inflationary observables perfectly match Planck data.
\end{abstract}


\maketitle


\section{Introduction}

After more than $30$ years since the idea of inflation was born\cite{Guth:1980zm}, inflation has become nowadays one of the cornerstones of  modern cosmology together with General Relativity and the Hot Big Bang Model. Today, no viable cosmological model can be constructed without an inflationary period.
Originally, inflation was proposed to solve the magnetic monopole problem \cite{Guth:1979bh}, but shortly afterwards it was realized that it can address the horizon and flatness problems of the standard Big-Bang cosmology as well.
It was also soon undertood that the density perturbations of the inflaton (typically a scalar field) can be the natural source of seeds needed to explain the onset of all the structures of the present universe \cite{Mukhanov:1981xt,Mukhanov:1982nu}.
Nowadays, it is the common lore to consider the inflaton as the main origin of density perturbations.

Meanwhile, since its discovery \cite{Penzias:1965wn}, the cosmic microwave background radiation (CMBR) has become a very powerful probe of the density power spectrum of our universe, and cosmology has evolved into a precision science as a result.
The detector sensitivity of CMBR experiments has been becoming better and better, and a variety of inflationary models have been excluded due to its power spectrum being inconsistent with observations.

At the theory side, precision observables of CMBR have been cornering the market of inflation model-building.
Most of inflationary models can be grouped into two categories: Small-field and large-field inflation.
In order to reproduce the observed scale invariant density power spectrum \cite{Ade:2013zuv}, inflaton is supposed to get through a phase of slow-roll along a flat enough potential.
However, small-field inflationary models typically suffer the so-called $\eta$-problem associated with a Hubble scale inflaton mass that causes a too strong red-tilte of power spectrum.
In the case of large-field inflation on the other hand, the accelerated expansion phase takes place when the inflaton has a field value larger than Planck scale (or the cut off scale of the given theory).
Such a trans-Planckian field value is not consistent with the concept of the effective field theory which is adapted in nearly all inflation set up's.
The point is that, in order to have a flat enough inflaton potential (required for a power spectrum consistent with observation) with such a large field value, one would need a severe fine tuning in order to suppress the effects of an infinite number of higher order operators, unless there is a very good symmetry which is not known or not justified yet.

If we stick to effective field theory approach for inflation model-building, it may be better to construct a model of small-field inflation. But small field inflation has its issues as well. Recently the BICEP2 experiment announced the discovery of tensor modes, \ie observations consistent with a large tensor-to-scalar ratio ($r\sim0.1$) \cite{Ade:2014xna}.
Such a claim is yet to be confirmed, nevertheless, if it turns out to be true, small-field inflation models are likely to be excluded  courtesy of the Lyth bound \cite{Lyth:1996im}.
However, one should be careful in interpreting the Lyth bound.
If the field space of the inflaton potential is one-dimensional, the Lyth bound is difficult to avoid, unless the shape of the potential is not simple \cite{Antusch:2014cpa}.
However, quite generically high energy theories have multi-dimensional scalar field spaces, and inflation can take place along a flat trajectory in such a space.
Hence, in such a case, there is a chance to avoid Lyth bound even if the field space for the inflation is limited to be sub-Planckian (See, 
for example, \cite{Silverstein:2008sg,McAllister:2008hb,Berg:2009tg,McDonald:2014oza,McDonald:2014nqa,Li:2014vpa}).

In this letter, we propose a novel scenario of a small-field inflation where the
$\eta$-problem is absent and a large tensor-to-scalar ratio of $r\sim0.1$ can be obtained without fine tuning even though the model belongs to the small field family, \ie it is subplanckian all along. All the remaining inflationary observables match perfectly well the observation of Planck satellite. 

\section{The model}
\begin{figure}[ht]
\begin{center}
\includegraphics[width=0.45\textwidth]{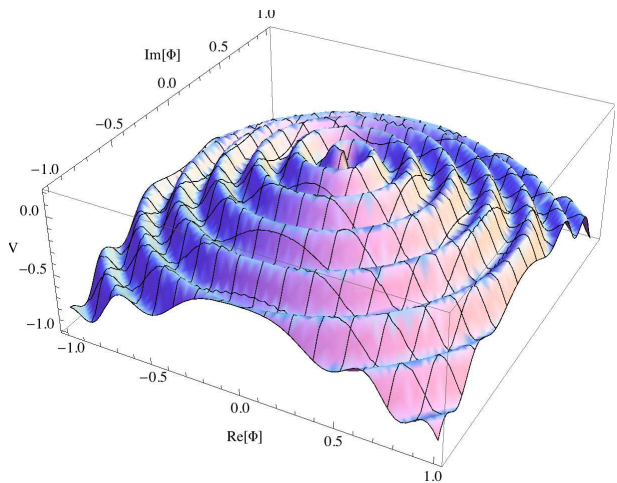}
\caption{Illustration of the potential in \eq{V} for $\Lambda^4 = 0.1 m^2 M_{\rm P}^2$, $M = 0.2 M_{\rm P}$, $\alpha=1$, and $\lambda=0$.}
\label{fig:V}
\end{center}
\end{figure}

We consider a potential \footnote{A similar type of potential has been considered at Ref.~\cite{Berg:2009tg,McDonald:2014oza,McDonald:2014nqa}. While they can be regarded as a sub-Planckian realization of chaotic inflation \cite{Linde:1986fd}, our scenario is a type of ``new inflation'' \cite{Linde:1981mu,Albrecht:1982wi} in a sub-Planckian scale.}:
\beq \label{V}
V = V_0 - m^2 |\Phi|^2 + \Lambda^4 \l[ 1 - \sin(2 \sqrt{2} \pi |\Phi|/M + \alpha \theta) \r] + \lambda |\Phi|^4
\eeq
where $\Lambda$ and $M$ are mass scales that will be constrained by inflationary phenomenology, $\alpha$ is a numerical constant, and $\theta = {\rm Arg}(\Phi)$.
Note that in \eq{V} the last term is added to stabilize $\Phi$.
This potential can be regarded as a simple example adapted to  ilustrate our idea.
In reality, $\Phi$ can be a Planck scale modulus, and its stabilization can be achieved in some other way rather than the $\lambda$-term here.
We assume $V_0 \gg \Lambda^4$.
Then, ignoring the terms in $\l[ ... \r]$ for a while and using $\Phi = \phi e^{i \theta}/\sqrt{2}$, one finds the minimum located at
\beq
\phi_0 = \frac{m}{\sqrt{\lambda}}
\eeq
and $V_0 = m^2 \phi_0^2/ 4$
for vanishing cosmological constant.
If inflation is driven by $V_0$, one finds
\beq \label{H-m}
m \approx 2 \sqrt{3} H_I \l( \frac{M_{\rm P}}{\phi_0} \r)
\eeq
where $H_I$ is the expansion rate during inflation. 
Hence, for $H_I \lesssim 10^{14} \GeV$, we need 
\beq
\lambda = \l( \frac{m}{\phi_0} \r)^2 \approx \l( \frac{2 \sqrt{3} H_I}{\phi_0} \frac{M_{\rm P}}{\phi_0} \r)^2 \lesssim 1.4 \times 10^{-8}
\eeq
where we used $\phi_0=M_{\rm P}$ for the inequality in the right-hand side.
The existence of the second term in $\l[ ... \r]$ of \eq{V} gives modulations to a simple tachyonic potential.
However, the important point is that there is a linear dependence on $|\phi|$ in the phase of the second term \footnote{A higher order $\phi$-dependence would work equally well, but, as can be seen in the discussion of the next section, a tuning would become necessary since, for a given $M$, the total change of $\phi$ for about $60$ $e$-foldings becomes smaller.
This tuning might be ameliorated by adjusting $M$ close to $\phi_0$ though.}.
So, depending on $\Lambda$, $M$ and $\alpha$, a spiral valley in the potential can exist as shown in Fig.~\ref{fig:V}.
This means that, if $\Phi$ is released from around the top of the potential, it is possible for $\Phi$ to follow the valley.
Note that, in this case, even if the radial direction of $\Phi$ is steep, the potential along the valley can be flat enough to pave the way for slow-roll inflation, depending on the potential parameters.
This is the subject of the next section.

\section{Inflation}
We set $\alpha=1$ to have a smoothly connected valley with the largest number of allowed turns.
The existence of a spiral valley means that along the $\phi$-direction a minimum appears almost periodically at least for a range of $\phi$ values, and inflaton may trace closely the minimum of the valley.
At the minimum, $\partial V/ \partial \phi = 0$ gives a relation
\beq
\frac{2 \pi \Lambda^4}{\phi M} \cos \theta_\phi = m^2 - \lambda \phi^2
\eeq
Note that inflation ends when this equality can not be satisfied, \ie  when $\phi$ becomes too large, satisfying
\beq
\phi \gtrsim \frac{2 \pi \Lambda^4}{m^2 M}
\eeq
where $-\lambda \phi^2 M$ term in the denominator was ignored.

The elements of the mass matrix at the minimum of the valley are given by:
\bea
\frac{\partial^2 V}{\partial \phi^2} &=& \frac{\Lambda^4}{M^2} ( 2 \pi )^2 \sin \theta_\phi - \l( m^2 - 3 \lambda \phi^2 \r)
\\
\frac{\partial^2 V}{\phi \partial \theta \partial \phi} &=& \frac{\Lambda^4}{\phi M} 2 \pi \sin \theta_\phi
\\
\frac{\partial^2 V}{\phi^2 \partial \theta^2} &=& \frac{\Lambda^4}{\phi^2} \sin \theta_\phi
\eea
Then, defining
\beq \label{a-delta}
a \equiv 2 \pi \frac{\phi}{M}, \quad \delta \equiv \frac{\l( m^2 - 3 \lambda \phi^2 \r) \phi M}{2 \pi \Lambda^4 \sin \theta_\phi}
\eeq
one finds the mass eigenvalues expressed as
\bea \label{mIsq}
m_\parallel^2 &=& - \l( \frac{\delta/a}{a - \delta + 1/a} \r) \frac{\Lambda^4}{\phi M} 2 \pi \sin \theta_\phi 
\\
m_\perp^2 &=& \l( a - \delta + \frac{1}{a} \r) \frac{\Lambda^4}{\phi M} 2 \pi \sin \theta_\phi 
\eea 
respectively. The inflaton is likely to follow the tachyonic direction.
Hence, the inflaton direction can be expressed as 
\beq
dI = \frac{\partial I}{\partial \phi} d\phi + \frac{\partial I}{\partial \theta} d\theta
\eeq
where
\bea
\frac{\partial I}{\partial \phi} &=& \frac{1}{\mathcal{N} a} \l[ 1 + \frac{\delta}{\l( a - \delta + 1/a \r)} \r]
\\
\frac{\partial I}{\partial \theta} &=& - \frac{\phi}{\mathcal{N}}
\eea 
with
\beq
\mathcal{N} = \sqrt{1 + a^{-2} \l[ 1 + \frac{\delta}{\l( a - \delta + 1/a \r)} \r]^2 }
\eeq
The slow-roll parameters are calculated as
\bea \label{ep}
\sqrt{2 \epsilon} &=& \l| \frac{M_{\rm P}}{V} \frac{\partial V}{\partial I} \r| = \l| \frac{M_{\rm P}}{V} \l( \frac{\partial V}{\partial \phi}  \frac{\partial \phi}{\partial I} + \frac{\partial V}{\partial \theta}  \frac{\partial \theta}{\partial I} \r) \r|
\\ \label{eta}
\eta &\equiv& \frac{M_{\rm P}^2}{V} \frac{\partial^2 V}{\partial I^2} \approx \frac{m_\parallel^2}{3 H^2}
\eea
The angular motion of the inflaton pushes inflaton outward from the minimum of the valley, and the shift is bounded to have some effects, which we expect on general grounds to be small.
So, we keep $\partial V/ \partial \phi$ factor in \eq{ep}.
On the other hand, such a  deviation will not make a sizable change in $\eta$ as long as the inflaton is mainly from angular direction.
Hence, we take $\eta$ as the one associated with the tachyonic direction.

Even though inflation takes place in a $2$-dimensional field space, it behaves as in the single field case.
Hence, in terms of slow-roll parameters, the inflationary observables are given by
\bea \label{PR}
P_{\mathcal{R}} &=& \frac{1}{24 \pi^2} \frac{V}{\epsilon M_{\rm P}^4}
\\ \label{ns}
n_s &=& 1 - 6 \epsilon + 2 \eta
\\
r &=& 16 \epsilon
\eea
In the next section, the result of numerical analysis is provided.

\section{Numerical analysis}
As justified in the numerical analysis, for the $e$-foldings relevant to our universe, one finds $\delta \sim 1$ and $a \gg 1$ resulting in $\mathcal{N} \approx 1$.
In this case, from \eqss{a-delta}{mIsq}{H-m}, one see that 
\beq
\eta \sim - \frac{1}{\pi^2} \l( \frac{M}{\phi} \r)^2 \l( \frac{M_{\rm P}}{\phi_0} \r)^2
\eeq
Also, when the second term of \eq{ep} is dominant, $\epsilon$ can be approximated to 
\beq
\sqrt{2 \epsilon} \sim \frac{2}{\pi} \l( \frac{M}{\phi_0} \r) \l( \frac{M_{\rm P}}{\phi_0} \r)
\eeq  
Hence, $M$ is constrained as
\beq
M \approx 0.18 \l( \frac{\epsilon}{6.7 \times 10^{-3}} \r)^{1/2} \frac{\phi_0^2}{M_{\rm P}}
\eeq
$\Lambda$ is also constrained, but it is neither simple nor particularly enlightening to show how.

Keeping in mind this key fact, we set the model parameters as  
\bea \label{paraset1}
& \frac{m}{10^{14} \GeV} = 0.796 \times 2 \sqrt{3}, \,\, \frac{\Lambda^4}{m^2 M_{\rm P}^2} = 4 \times 10^{-3} & 
\\ \label{paraset2}
& \frac{M}{M_{\rm P}} = 7.3 \times 10^{-2}, \,\, \lambda = (m/M_{\rm P})^2
\eea
for the numerical analysis.
This set of values (which by no means is unique and has to be regarded as one point in a region of parameters space) is just an example for which we can obtain a result matching perfectly the observation of Planck satellite \cite{Ade:2013zuv}. 
%
\begin{figure}[ht]
\begin{center}
\includegraphics[width=0.45\textwidth]{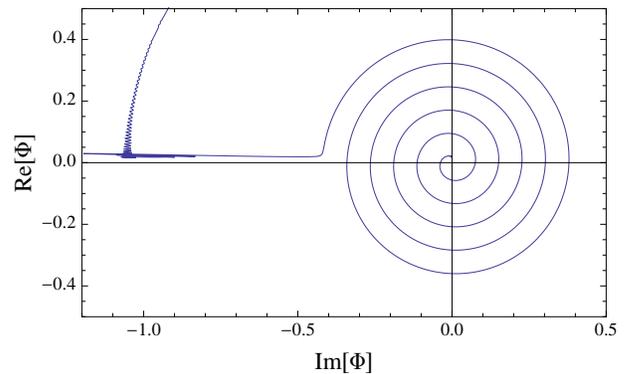}
\caption{Trajectory of inflaton}
\label{fig:trajectory}
\end{center}
\end{figure}
%
\begin{figure}[h!]
\centering
\includegraphics[width=0.45\textwidth]{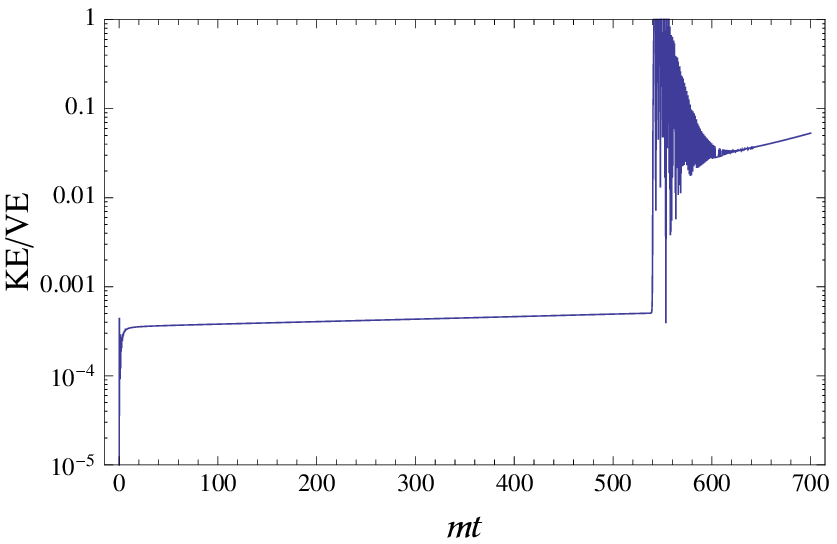}
\includegraphics[width=0.45\textwidth]{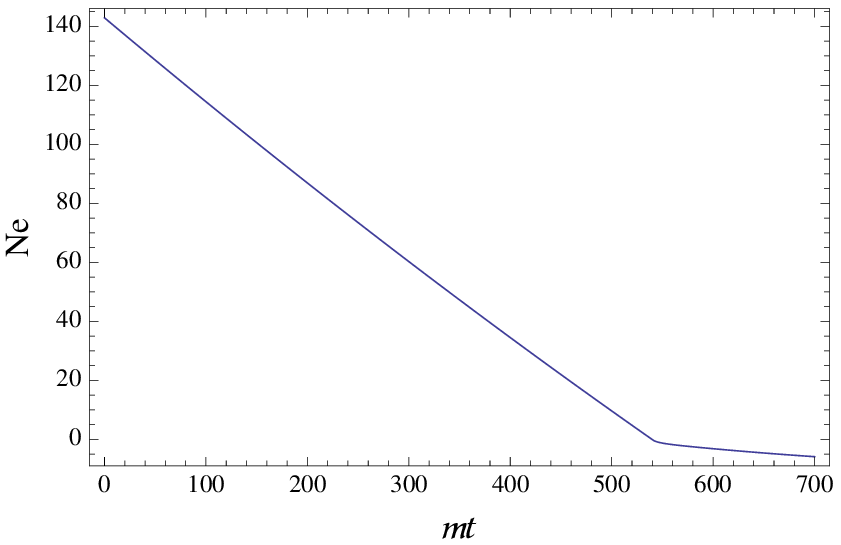}
\caption{\label{fig:inflation} \textit{Upper}: The ratio of kinetic energy to the potential. 
\textit{Lower}: $e$-foldings. The appearance of negative value is because the convention of $N_e$ calculated from the end of inflation.}
\end{figure}
In the field basis, the equations of motions are given by
\bea
0 &=& \ddot{\phi} + 3 H \dot{\phi} + \frac{\partial V}{\partial \phi}
\\
0 &=& \ddot{\theta} + \frac{\dot{\phi} \dot{\theta}}{\phi} + 3 H \dot{\theta} + \frac{\partial V}{\phi^2 \partial \theta}
\eea
Integrating these equations with $\phi_i=0.3 M$ as the initial field value, we obtain an inflaton trajectory following the spiral valley of the potential, as shown in Fig.~\ref{fig:trajectory}.
Note that, as can be seen from the figure, inflation takes place during the spiral motion and ends once the minimum along $\phi$ disappears or becomes too shallow to keep holding  inflaton along the valley.
The number of $e$-foldings is large enough, as shown in bottom plot of Fig.~\ref{fig:inflation}, even though $\phi$ is well below Planck scale during inflation.
%
\begin{figure}[h!]
\centering
\includegraphics[width=0.45\textwidth]{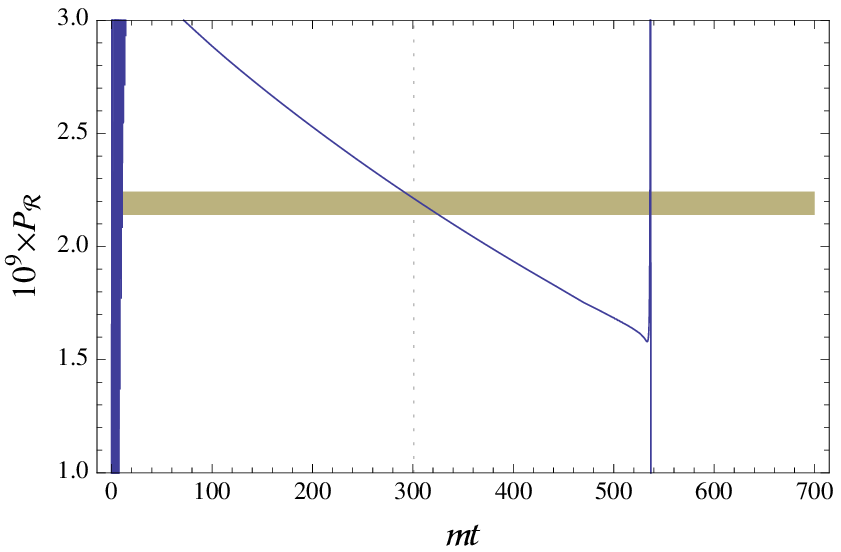}
\includegraphics[width=0.45\textwidth]{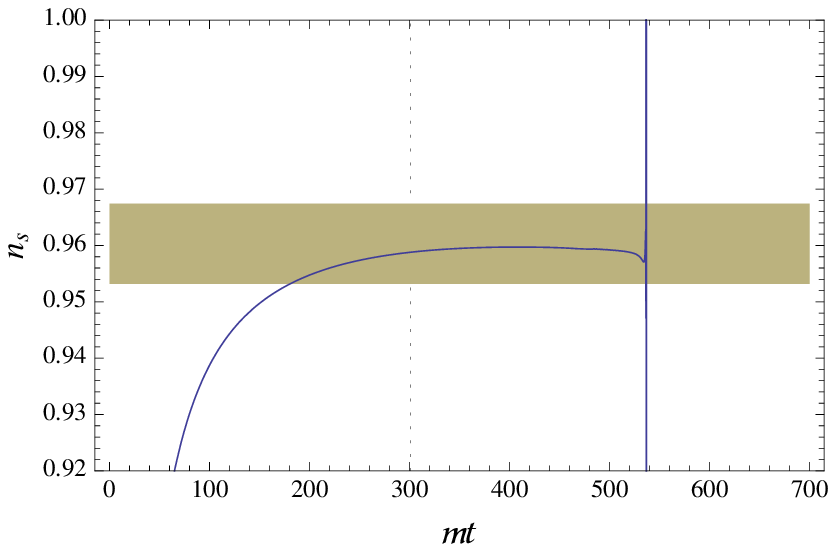}
\includegraphics[width=0.45\textwidth]{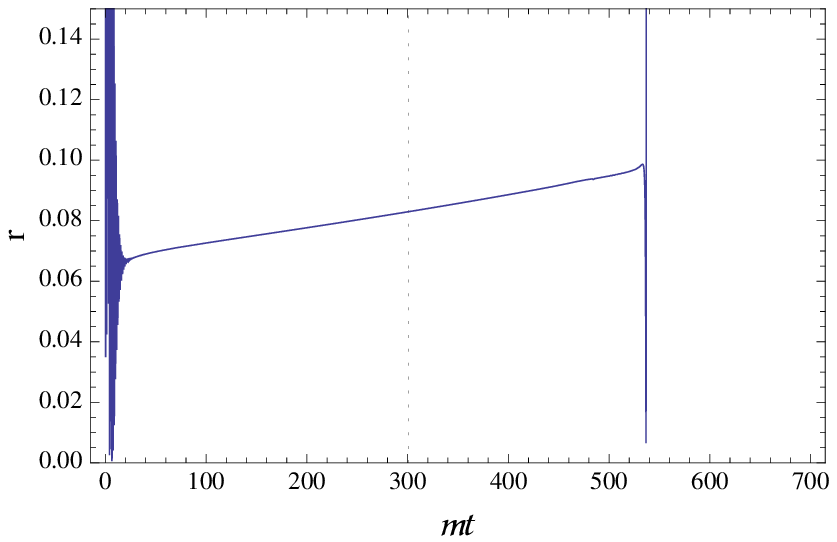}
\caption{Inflationary observables.
The horizontal color band is the uncertainty of Planck data at $68$ \% CL, and the vertical dotted line corresponds to $N_e=60$. 
The initial rapid oscillatory behavior is due to the initial position of field that is deviated slightly from the minimum of valley.}
\label{fig:observables}
\end{figure}
The observables of inflation are shown in Fig.~\ref{fig:observables}, 
and as can be seen there this model perfectly agrees with the observations of Planck satellite.
We also found that the running of the spectral index is of $\mathcal{O}(10^{-4})$ for most of $60$-efoldings. Therefore spiral inflation predicts a running that is so small that it is essentially experimentally indistinguishable from zero running.
%
Although it is not the only exception found, it is very interesting to notice that unlike most small single field models of inflation, spiral inflation  can provide a large tensor-to-scalar ratio of $\mathcal{O}(0.1)$ as claimed recently by the BICEP2 experiment. 
This result (if taken face value) seems to contradict  Lyth bound \cite{Lyth:1996im}, but actually it is not since the length of the trajectory of our inflaton is actually longer than Planck scale, but it is curled up within a scale smaller than Planck scale. 

A remark is in order before we conclude.
In our scenario, it is mandatory that, as the initial condition, $\Phi$ had to be around the top of the potential before the beginning of the spiral inflation.
This might be achieved by, for example, a stage of thermal inflation \cite{Lyth:1995hj,Lyth:1995ka} during which thermal effects holds $\Phi$ around the origin.
Also, compared to the sub-Planckian realizations of chaotic inflation discussed in \cite{Berg:2009tg,McDonald:2014oza,McDonald:2014nqa}, our scenario provides a natural set up for a (relatively) low reheating temperature which is useful to avoid the possible repopulation of dangerous relics.

\section{Conclusions}

In this paper, we proposed a novel scenario of a small field inflation which is free from the long standing $\eta$-problem, which plagues the sub-Planckian inflationary models, without fine tuning. 
Very interestingly, while fully consistent with the observation of Planck satellite, it provides easily a large tensor-to-scalar ratio, $r \sim 0.1$ which apparently has been observed recently by BICEP2 experiment.
Motivated by the nice features presented by spiral inflation, it would be very interesting to search a full UV realization of our model.

\section*{Acknowledgements}
The authors acknowledge support from the MEC and FEDER (EC) Grants FPA2011-23596 and the Generalitat Valenciana under grant PROMETEOII/2013/017.
G.B. acknowledges partial support from the European Union FP7 ITN INVISIBLES (Marie Curie Actions, PITN-GA-2011-289442).


\end{document}